\documentclass[aps,prd,twocolumn,groupedaddress,showpacs]{revtex4-1}

\usepackage{graphicx}  
\usepackage{aas_macros}

\begin{document}

\title{Three-dimensional Hydrodynamic Simulations \\
of the Combustion of a Neutron Star into a Quark Star} 
\author{Matthias Herzog} 
\affiliation{Max-Planck-Institut f{\"u}r Astrophysik,
  Karl-Schwarzschild-Str.~1, D-85741 Garching, Germany}
\email{mherzog@mpa-garching.mpg.de}
\author{Friedrich K. R\"opke} 
\affiliation{Institut f{\"u}r Theoretische Physik und Astrophysik, Universit\"at W\"urzburg,
  Emil-Fischer-Str.~31, D-97074 W{\"u}rzburg, Germany}

\date{\today}

\begin{abstract}

We present three-dimensional numerical simulations of turbulent
combustion converting a neutron star into a quark star. Hadronic
matter, described by a micro-physical finite-temperature equation of
state, is converted into strange quark matter. We assume this phase,
represented by a bag-model equation of state, to be absolutely
stable. Following the example of thermonuclear burning in white dwarfs
leading to Type Ia supernovae, we treat the conversion process as a
potentially turbulent deflagration. Solving the non-relativistic Euler
equations using established numerical methods we conduct large eddy
simulations including an elaborate subgrid scale model, while the
propagation of the conversion front is modeled with a level-set
method.  Our results show that for large parts of the parameter space
the conversion becomes turbulent and therefore significantly faster
than in the laminar case. Despite assuming absolutely stable strange
quark matter, in our hydrodynamic approximation an outer layer remains
in the hadronic phase, because the conversion front stops when it
reaches conditions under which the combustion is no longer exothermic.

\end{abstract}
\maketitle

\section{Introduction}

Based on earlier work by \citet{bodmer1971a} and \citet{itoh1970a},
\citet{witten1984a} suggested that a mixture of about the same number
of \mbox{u-,} d- and s-quarks, called strange quark matter (SQM), was
the true ground state of matter, whereas ordinary nuclear matter is
only a metastable, yet usually extremely long-lived state. This
conjecture, known today as \textit{strange matter hypothesis}, was
discussed lively ever since, but no final verdict about its
correctness could be made because the equation of state (EoS) of cold
dense matter is still largely unknown. Matter in this extreme state is
inaccessible to laboratory experiments; compact stars, however, offer
a possibility to test the strange matter hypothesis.
Shortly after Witten's work also \citet{haensel1986a} and
\citet{alcock1986a} proposed \textit{strange stars}, compact stars
consisting entirely of SQM.
\citet{alcock1986a} based their work on
the idea that compact stars are not born as
strange stars, but as hadronic neutron stars, which later are
converted into strange stars or \textit{hybrid stars} -- compact stars
consisting of a quark core and hadronic outer layers.

Hadronic matter does not decay into SQM spontaneously, even though it
would be energetically favorable, because this process would require
a large amount of simultaneous weak reactions -- the probability for
this to happen is vanishingly low.  But if some SQM already exists
inside a neutron star, the diffusion of s-quarks from this seed into
the surrounding hadronic matter would convert it into SQM.  This
conversion process should take place in a confined region and on
length-scales small compared to the size of the star. It is expected
to occur only if the conversion releases energy, that is, if it is an
exothermic process.  The described situation is therefore similar to
the propagation of a chemical flame, or even more similar to the
thermonuclear burning inside a white dwarf during a Type Ia supernova
(SN~Ia).  Thus, it is natural to think of the conversion of hadronic
matter into SQM as a ``combustion''. In the spirit of this analogy, we
will sometimes refer to the conversion as ``burning'' and to the
conversion front as ``flame front''.
\citet{alcock1986a} were the first to suggest that a strange
star may originate from a combustion of an ordinary neutron star. They
also considered how a SQM seed which subsequently triggers the
conversion into a strange star may come about and described various
possibilities by either internal nucleation or external
seeding. Subsequently the idea of a combustion was discussed in more
detail by various authors \citep{horvath1988a,olesen1991a,cho1994a,lugones1994a,lugones1995a,
 tokareva2006a,drago2007a,niebergal2010a}.

The laminar conversion velocity was first estimated by
\citet{olinto1987a}, and, with similar results, by
\citet{heiselberg1991a}. Based on their results, \citet{olesen1991a}
calculated the burning of a neutron star using a one-dimensional model
with laminar burning and obtained conversion timescales from
$10^{-1}\, \mathrm{s}$ to $10^{2}\, \mathrm{s}$.
 \citet{horvath1988a} suggested that the combustion
should be turbulent due to various instabilities of the conversion front and
therefore the conversion velocity should be enhanced considerably (see
\cite{horvath2010a} for a recent update).
\citet{lugones1994a} and \citet{lugones1995a} pointed out the importance of the conditions for
an exothermic combustion.  The combustion mode was discussed from a
hydrodynamic point of view also by \citet{cho1994a},
\citet{tokareva2006a} and \citet{drago2007a}, where the latter
expected the burning to be subsonic, although accelerated by
turbulence.
New ideas concerning the initial seeding were recently published by
\citet{perez-garcia2010a}. They suggested that the self-annihilation
of weakly interacting dark matter particles (WIMPs) inside a neutron
star may provide a SQM seed.
Recently, hydrodynamic simulations of the combustion front were presented
by \citet{niebergal2010a}. Their results of the laminar conversion
velocity differed strongly from earlier estimates.
On the observational side \citet{leahy2008a}, extended in
\citet{ouyed2010a}, examined the supernova SN~2006gy and suggested
that this extremely luminous event can be explained by a ``quark
nova'' -- the transition of the newly formed neutron star to a strange
quark star shortly after a core collapse supernova of a very massive
star.

Here we study the dynamical behavior of the conversion inside a
neutron star. We model the conversion as a combustion, particularly as
a subsonic deflagration. As mentioned above, it is widely assumed that the
conversion process turns turbulent \citep[e.g.][]{drago2007a,
  horvath1988a, horvath2010a}, but dynamical, multi-dimensional
simulations have never been performed. Thus, our main focus will be to
explore if and how turbulent motion occurs during the conversion
process and to which consequences for the final state of the neutron
star this may lead.

This work is organized as follows: In Section \ref{sec:eos} we
describe the EoS that we use in our calculations. In
Section \ref{sec:combustion} we introduce our concept of modeling the
conversion as a turbulent combustion, and in Section
\ref{sec:numerical_method} our numerical method is explained. We
present numerical simulations and their results in Section \ref{sec:simulations}
and conclude in Section \ref{sec:conclusions}.

\section{Equation of State}\label{sec:eos}

\subsection{Equation of State for Hadronic Matter}

We consider the two micro-physical, finite temperature EoS which are
most frequently used in simulations of astrophysical events such as
core collapse supernovae and neutron star mergers: the EoS by
\citet{lattimer1991a} (LS EoS) and by \citet{shen1998a} (Shen
EoS). The LS EoS is based on a non-relativistic liquid drop model with
an incompressibility modulus of $K=180\,\mathrm{MeV}$. For calculating
the Shen EoS relativistic mean field theory was applied, here
$K=280\,\mathrm{MeV}$ is adopted. 

The recent measurement of the Shapiro delay of the binary millisecond
pulsar \mbox{J1614-2230} \cite{demorest2010a} yields a gravitational
mass of the pulsar of $M=(1.97\pm0.04)\, M_\odot$.  In contrast to the
Shen EoS, the LS EoS is rather soft. Consequently it leads to a
maximum mass for a hadronic non-rotating neutron star of only
$M_\mathrm{max}^\mathrm{LS}\sim 1.8 M_\odot$ and is therefore in
conflict with the observation of pulsar \mbox{J1614-2230}. We
nevertheless use the LS EoS in this work, because we do not claim to
conduct realistic simulations but we rather see our work as a first
step into this so far mostly unexplored field. An alternative would be
to change the incompressibility modulus of the LS EoS to
$K=220\,\mathrm{MeV}$, which leads to a maximum mass compatible with
the observations.
We discuss this possibility briefly in Section \ref{sec:exo}.

For simplicity we assume for all our calculations a constant low
proton fraction $Y_p$. Variations of its value, particularly assuming
$\beta$-equilibrium, do not lead to a significant change of our
results, as is shown exemplary in Section \ref{sec:exo}.
In the same section we explain that for physical reasons it turned out
that it was impossible to use the Shen EoS, thus we perform all our
simulations using the LS EoS.

\subsection{Equation of State for Strange Quark
  Matter}\label{sec:quarkeos}

We describe SQM by a simple bag model for finite temperatures
\citep{cleymans1986a} based on the MIT bag model \citep{chodos1974a}.
This model treats SQM as ideal Fermi gases of massless and
non-interacting u-, d-, and s-quarks inside a confining bag. Since in
this approximation the quarks can be described by only one chemical
potential, the resulting analytic expressions for pressure $P$, energy
density $e$ and baryon number density $n$ as function of the bag
constant $B$ and the chemical potential $\mu$ are
\citep{madsen1999a,tokareva2006a}
\begin{eqnarray}
P&=& \frac{19}{36}\pi^2T^4+\frac{3}{2}T^2\mu^2+\frac{3}{4\pi^2}\mu^4-B,\\
e&=& \frac{19}{12}\pi^2T^4+\frac{9}{2}T^2\mu^2+\frac{9}{4\pi^2}\mu^4+B,\\
n&=& T^2\mu+\frac{1}{\pi^2}\mu^3,
\end{eqnarray}
which corresponds to the simple pressure-density relation
\begin{eqnarray}
P &=& \frac{1}{3}(e-4B). \label{sqmeos}
\end{eqnarray}

The value of $B$ is not known; however, some constraints can be
derived. We can specify a lower limit of $B$ due to the fact that
nucleons do not decay spontaneously to two-flavor quark
matter. \citet{madsen1999a} shows that this lower limit is $ B^{1/4}
\ge 145\, \mathrm{MeV}$ and gives an expression for the energy
per baryon $E/A$ as function of $B$,
\begin{eqnarray}
E/A &=& 829\, \mathrm{MeV} \frac{B^{1/4}}{145\, \mathrm{MeV}}. \label{EA}
\end{eqnarray}
Since nuclear matter has an energy per baryon of $E/A \sim 930\,
\mathrm{MeV}$, according to (\ref{EA}) bag constants lower
than $B^{1/4} = 160\, \mathrm{MeV}$ correspond to absolutely stable SQM.

The next step to a more realistic EoS would be to include the masses
of the quarks. Although the current masses of u- and d-quarks are at
most $10\, \mathrm{MeV}$ and are therefore negligible, the mass of the
s-quark is of the order of $100\, \mathrm{MeV}$. However, in this case
an analytic expression for $P$, $e$ and $n$ is no longer possible for
finite temperatures. Including quark masses as well as QCD
interactions \citep{farhi1984a} leads, for example, at a given $B$ to
a the energy per baryon which is about $20\, \mathrm{MeV}$
higher than given by (\ref{EA})\citep{bauswein2010a} and thus shifts
the range of bag constants in which SQM is absolute stable.

\section{Combustion}\label{sec:combustion}

We model the conversion from hadronic matter into SQM as a combustion,
initiated by a seeding of SQM which we assume to occur in the center
of the star.  We do not specify the origin of the initial SQM seed
(see \citep{alcock1986a} for various possibilities, or
\citep{perez-garcia2010a} for new ideas). The flame front, initially
consisting of the boundary surface of some central seed, propagates
outwards and converts hadronic matter into SQM, provided this reaction is
exothermic.
If this is the case, the difference in the energy per baryon is
released into internal energy and therefore the temperature increases.
The analogous case in chemical combustion theory is called premixed
combustion, where fuel and oxidizer are already mixed at low
temperatures and the flame propagates by conduction of heat
\citep{peters2000a}.  In the case of the burning of hadronic matter
into SQM the abundance of s-quarks plays the role of temperature;
accordingly the diffusion of s-quarks leads to the propagation of the
flame front.
The combustion process takes place on length scales of the
micro-physical reactions, which can be estimated as follows: The
disintegration of a nucleon into quarks happens on time scales of the
strong interaction, $\sim 10^{-24}\, \mathrm{s}$, corresponding to a
length scale of $\sim 10^{-13}\, \mathrm{cm}$. The conversion of a
d-quark into an s-quark due to the weak interaction takes place in $\sim
10^{-8}\, \mathrm{s}$. Since the weak processes are much slower, they
determine the time scale of the burning, leading to a width of the
reaction zone, $l_\mathrm{burn}$, not exceeding $10^{2}\,
\mathrm{cm}$, whereas realistic calculations yield
$l_\mathrm{burn}\sim 10\, \mathrm{cm}$ \citep{niebergal2010a}. These
length scales are much smaller than the resolution we can achieve in
our simulations ($l_\mathrm{resolved}> 10^{3}\, \mathrm{cm}$) and
therefore we cannot resolve the reaction zone. Instead, we model the
conversion front as a discontinuity which separates the ``unburnt''
(hadronic) matter from the ``burnt'' (strange quark) matter and have
to take the propagation velocity of the conversion front with respect
to the fluid flow as an input parameter, since this velocity is not
determined by the hydrodynamic equations but by micro-physical
processes on scales of the internal structure of the conversion front.

A combustion can take place either as a supersonic detonation driven
by a shock wave, or as a subsonic deflagration driven by diffusion
processes. Since we cannot resolve the internal structure of the flame
we have to decide before starting our computations whether to model
the conversion as a deflagration or as a detonation.
\citet{drago2007a} examine the conversion of hadronic matter into quark
matter based on the hydrodynamic jump conditions. They assume the
combustion to start as a deflagration and conclude that the process
should stay subsonic. Also \citet{niebergal2010a} and
\citet{horvath2010a} assume the conversion to be subsonic. Based on
these recent publications we decided to choose a deflagration as
combustion mode, though we do not exclude the detonation mode and
might consider it in future work.

The relevant input velocity for a deflagration is the
laminar burning velocity $v_\mathrm{lam}$, which is only very poorly
known for the burning of hadronic matter into SQM. The first attempts to
determine it where made by \citet{olinto1987a}, who estimates
$v_\mathrm{lam}$ based on the diffusion of strange quarks and the
equilibration of the SQM via weak interactions. The resulting
velocities are generally rather low but strongly temperature dependent
and would lead to a wide range of neutron star conversion timescales
from milliseconds up to several minutes.  Recently,
\citet{niebergal2010a} conducted one-dimensional hydrodynamic
simulations of the combustion flame, including neutrino emission and
strange quark diffusion. They found laminar burning velocities much
higher than in earlier work. Because the methods of
\citet{niebergal2010a} are more sophisticated than in previous
publications, we adopt a weakly density-dependent laminar burning
velocity based on a linear fit to their results. This leads to
$v_{lam}\sim10^8\, \mathrm{cm/s}$ in the center of the initial neutron
star at densities of $e\sim10^{15}\, \mathrm{g/cm^3}$. Since according
to our simulations the burning velocity is strongly enhanced by
turbulence, the importance to know the exact value of $v_{lam}$ is
rather subordinate (see below).

\subsection{Turbulent Combustion}\label{sec:turb_comb}

Under certain conditions the laminar propagation of the conversion
front can be distorted by Rayleigh-Taylor (buoyancy) instabilities
\citep[see][and references therein]{timmes1992a}. A necessary
condition for this is that the gradient of the gravitational potential
and the gradient of the total energy density point in opposite
directions (``inverse density stratification'').

In chemical flames, as well as during the thermonuclear burning of
carbon and oxygen in the center of a white dwarf, the large amount of
energy released during the burning process leads to a sharp increase
in temperature. 
In chemical flames a strong increase of pressure, or a strong decrease
in density at constant pressure, is the natural result and therefore
is usually taken for granted in qualitative considerations. Similarly,
in SNe~Ia the degeneracy of the matter is partially lifted, therefore
the density decreases also in this case, albeit not as strongly as in
chemical flames.  Moreover, in these cases, although the chemical
abundances change during the burning process, the EoS does not change
dramatically. In the case of the burning in white dwarfs at densities
$\lesssim 7-8\times10^{9}\mathrm{g/cm^3}$ this leads to an inverse
density stratification, instabilities and turbulence
\citep{timmes1992a}.
However, because of the strongly degenerate state of matter in neutron
stars and the fundamentally different EoS before and after the
conversion process it cannot be taken for granted that the neutron
star matter behaves in the same way as described above. The state of
the fluid behind the conversion front is determined by the change of
the EoS and the hydrodynamic jump conditions \citep[see
  e.g.][]{drago2007a} which result from the conservation of the baryon
flux density and the energy-momentum tensor across the flame surface
and has to be computed in hydrodynamic simulations.
To explore if in the vicinity of the propagation front the density of
the SQM is lower than the density of the hadronic phase for our choice
of EoS is therefore one aim of this work. 

The Rayleigh-Taylor instability can only grow and lead to turbulent
motion if the perturbations of the front exceed some minimal length
scale, $\lambda_\mathrm{min}$, which depends on the burning velocity,
the gravitational acceleration $g$, and the density contrast between the
total energy density of the hadronic phase $e_h$ and the total
energy density of the quark phase $e_q$ \citep{timmes1992a},
\begin{eqnarray}
  \lambda_\mathrm{min}&=&2\pi
  v^2_\mathrm{lam}\left(g\frac{e_h-e_q}{e_h+e_q}\right)^{-1}. \label{eq:lambdamin}
\end{eqnarray}
We calculate $\lambda_\mathrm{min}$ for different setups in Section
\ref{sec:turb}.

In the established heuristic turbulence model
\citep{richardson1922a,kolmogorov1941a} instabilities like the
Rayleigh-Taylor instability (and secondary shear instabilities) lead
to turbulent eddies on large scales, which decay successively into
ever smaller eddies until, at the Kolmogorov length scale $l_K$,
viscosity effects dissipate the smallest eddies into thermal energy.
In this \textit{turbulent cascade} kinetic energy is transported from
the largest to the smallest scales and is finally dissipated. This
picture assumes that magnetic fields do not significantly affect the dynamics.  For the
velocity fluctuation $v(l)$ on a given scale $l$, which can be
interpreted as the turnover velocity of an eddy of size $l$, this
model yields the \textit{Kolmogorov scaling}
\citep{landaulifshitz6eng},
\begin{eqnarray}
  v(l)&=&v(L)\left(\frac{l}{L}\right)^{1/3}, \label{eq:vofl}
\end{eqnarray}
where $L$ is the integral scale, the size of the largest eddies.

The Reynolds number $Re$ on different scales is therefore
\begin{eqnarray}
Re(l)&=& Re(L)\left(\frac{l}{L}\right)^{4/3},\label{eq:re}
\end{eqnarray}
since $Re(l)\propto v(l)l$. \citet{horvath1988a} estimate the Reynolds
number of flows in both neutron and strange stars to be $Re(L)\sim
10^{10}$. At the Kolmogorov scale $Re(l_K)\sim 1$ holds, so we get
\begin{eqnarray}
l_K &=&L\left(\frac{Re(l_K)}{Re(L)}\right)^{3/4}\sim10^{-8}\,\mathrm{cm},
\end{eqnarray}
and hence $l_K<<l_\mathrm{burn}$.

The scale on which the eddy turnover velocity is equal to the laminar
burning velocity is defined as the Gibson scale $l_G$
\citep[e.g.][]{peters2000a}, 
\begin{eqnarray}
  v(l_G)&=& v_\mathrm{lam}. \label{eq:gibson}
\end{eqnarray}

Turbulence cannot distort the flame front on scales smaller than $l_G$
since according to (\ref{eq:vofl}) on these scales the eddy turnover
velocity is smaller than the laminar burning velocity, whereas on
scales larger than $l_G$, the turbulent eddies alter the the shape
of the flame front.

If we assume the above scaling law for a rising Rayleigh-Taylor
unstable bubble of typical size $L\approx 10^{5}\, \mathrm{cm}$ and
typical macroscopic velocity variations $v(L)\approx 10^{9}\,
\mathrm{cm/s}$, we find $l_G = 10^2\, \mathrm{cm}$.  
The combustion theory was developed for chemical flames and only
adapted to SNe~Ia \citep{niemeyer1997b,niemeyer1997d}, whereas no
detailed studies were conducted for the case treated in this work.
However, the Kolmogorov scaling was found to fit quite well in the
case of SNe~Ia \citep{ciaraldi2009a, zingale2005a}, so based on these
results and in the absence of exact calculations we assume that this
is the case for our problem as well and obtain
\begin{eqnarray}
  l_\mathrm{burn} < l_G < l_\mathrm{resolved}.
\end{eqnarray}
This leads to two important consequences: $l_\mathrm{burn} < l_G$
means that the turbulent eddies cannot disturb the flame front. Thus,
it can still be described as a well-defined discontinuity.  The
burning is said to take place in the \textit{flamelet regime}
\citep{peters2000a}: Although the internal flame structure is not
disturbed, the total burning rate is enhanced as turbulence alters the
geometry and thus enlarges the surface of the front.  Since $l_G <
l_\mathrm{resolved} $ the surface of the flame front is also enhanced
on unresolved scales, leading to an increase in the effective front
propagation velocity on these scales.
This effective velocity is described by the \textit{turbulent burning
  velocity} $v_\mathrm{turb}$, which is defined as the mean
propagation velocity of the flame front at the marginally resolved
scale.

For strong turbulence, the turbulent burning velocity becomes
independent of the laminar burning velocity, as is the case during the
thermonuclear burning of a white dwarf. In this work we aim to explore
if the same is true in the conversion process of a neutron
star.

\subsection{Conditions for Exothermic Combustion}\label{sec:exo}
\begin{figure}
\includegraphics[width=8.6cm]{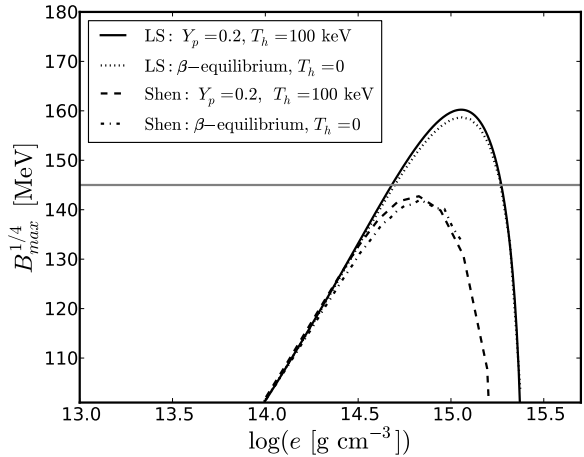}
\caption{Maximum bag constant $B$ allowing an exothermic combustion as
  function of the total energy density $e$ for two different hadronic
  EoS (Lattimer-Swesty with $K=180\,\mathrm{MeV}$ and Shen). The
  horizontal line indicates the theoretical lower limit of $B$. For
  each EoS, two cases are plotted: in the first case temperature $T_h$
  and proton fraction $Y_p$ are kept constant, in the second case the
  matter is in $\beta$-equilibrium at zero temperature.}
\label{fig:plotcomb}
\end{figure}

Since we describe the conversion of hadronic matter into SQM as a
combustion, and a combustion has to be, by definition, exothermic
\citep{anile1989a}, we can specify the following necessary condition
for the conversion to take place: The total energy density of
the quark phase $e_q$ in a thermodynamic state $(P,X)$ has to be
lower than the energy for the hadronic matter $e_h$ in the same state
\citep{anile1989a},
\begin{eqnarray}
e_h(P,X) \,& > &\, e_q(P,X)\label{exo},
\end{eqnarray}
where $P$ is the pressure, $X$ is the generalized volume,
$X=(e+P)/n^2_B$, and $n_B$ is the baryon density.  In the case of our
analytic EoS for SQM (\ref{sqmeos}), this can be rewritten as a simple
condition for the energy density of the hadronic phase
\citep{barz1985a,lugones1994a} :
\begin{eqnarray}
e_h(P) \,& > &\, 3P+4B.\label{eq:exo2}
\end{eqnarray}
From this relation it becomes clear that for each given total energy
density $e_h$ and temperature $T_h$ the corresponding pressure of the
hadronic phase $P$ and the value of $B$ determine whether the
conversion can proceed in form of a combustion wave. Thus, after
choosing the EoS and assuming a fixed $T_h$ we can calculate for each
$e_h$ a critical bag constant, $B_\mathrm{crit}(e_h)$, which is the
largest possible bag constant for an exothermic combustion. The
results of these calculations using both the LS EoS with
$K=180\,\mathrm{MeV}$ and the Shen EoS are shown in Figure
\ref{fig:plotcomb}.  Here the results are plotted for two different
cases: In the first case we assume a constant temperature of the
unburnt hadronic matter of $T_h=100\, \mathrm{keV}$ and a constant
proton fraction of $Y_p=0.2$. We adopt these assumptions for our
numerical simulations presented in Section \ref{sec:combustion}. In
the second case we assume $\beta$-equilibrium and zero temperature. As
visible in Figure \ref{fig:plotcomb}, the differences between the two
cases are rather small and thus negligible for the qualitative
treatment in this work.
Also apparent from this figure is that for bag constants larger than
the theoretical lower limit, $B^{1/4}>145\, \mathrm{MeV}$, and
temperatures found in the interior of cold neutron stars, hadronic
matter described by the Shen EoS cannot be burned into SQM in an
exothermic combustion, regardless of the density. In contrast, matter
described by the LS EoS can be converted into SQM in an exothermic way
at densities occurring in the center of neutron stars. 
The difference between the two hadronic EoS can be explained as
follows. The Shen EoS is rather stiff, much stiffer than the LS
EoS, that is at the same density the pressure is much higher. According to
(\ref{eq:exo2}) this leads to a higher energy threshold for a given density.
Based on this results we have to refrain from using the Shen EoS in
our hydrodynamic simulations.

\begin{figure}
\includegraphics[width=8.6cm]{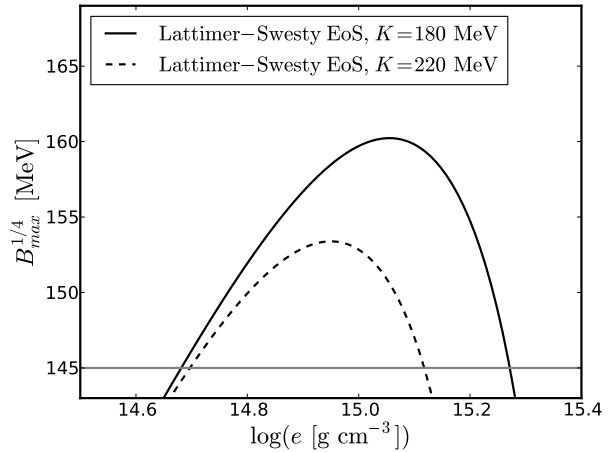}
\caption{Maximum bag constant $B$ allowing an exothermic combustion as
  function of the total energy density $e$ for the LS EoS and two
  different incompressibility moduli $K$. The horizontal line
  indicates the theoretical lower limit of $B$.}
\label{fig:plotcomb3}
\end{figure}

The LS EoS can be used with different incompressibility moduli $K$, we
consider $K=180\,\mathrm{MeV}$ and $K=220\,\mathrm{MeV}$. We compare
these two possibilities in Figure \ref{fig:plotcomb3}. For low bag
constants ($B^{1/4}\sim145\, \mathrm{MeV}$) the higher stiffness of
the EoS with higher $K$ affects the lower density limit only slightly,
but for $B^{1/4}\gtrsim152\, \mathrm{MeV}$ the range in which
exothermic combustion is possible becomes very narrow. 
Since our goal is to conduct simulations with higher bag constants to
be able to compare the results for a wide range in the amount of
released energy, we use in our simulations only the LS EoS
with $K=180\,\mathrm{MeV}$.
\begin{figure}
\includegraphics[width=8.6cm]{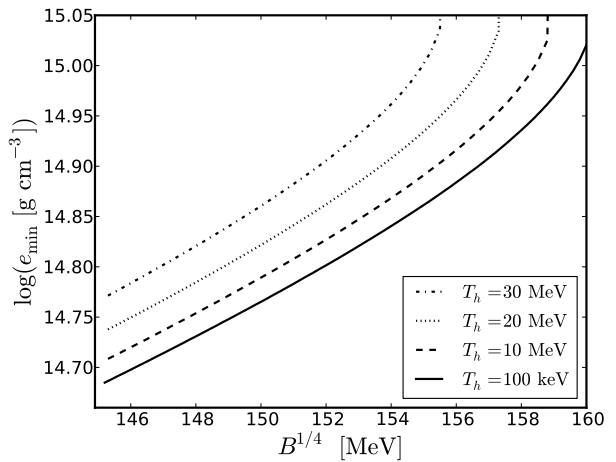}
\caption{Minimum total energy density $e$ for an exothermic combustion
  as function of the bag constant $B$ and for different temperatures
  $T_h$, using the Lattimer-Swesty EoS with $K=180\,\mathrm{MeV}$.}
\label{fig:plotcomb2}
\end{figure}
In Figure \ref{fig:plotcomb2} we concentrate on this case. Here we
plot the minimum total energy density of the hadronic phase,
$e_\mathrm{min}(B)$, as a function of $B$ and for different fixed
temperatures. Since below this density threshold no combustion is
possible, it plays an important role in our simulations.
The continuous line in Figure \ref{fig:plotcomb2} shows the case with
$T_h=100\, \mathrm{keV}$, the temperature we adopt for the cold
neutron star in our simulations.  In addition we explore the effects
of several higher temperatures. For temperatures up to $T_h=1\,
\mathrm{MeV}$ only slight differences would be visible due to the
strong degeneracy of the matter.  In proto-neutron stars considerably
higher temperatures occur, therefore also results for $T_h=10\,,20\,$
and $30\,\mathrm{MeV}$ are shown in the figure. These temperatures
have a noticeable effect on the density threshold, as visible in
Figure \ref{fig:plotcomb2}.  In general, higher temperatures move the
density threshold to higher densities.

\section{Numerical Method}\label{sec:numerical_method}

Taking advantage of the methodical similarities of combustion in white
dwarfs and in neutron stars, we use an existing code, which is well
tested and frequently applied for various SN~Ia  related simulations
\citep[e.g.][]{reinecke2002d,roepke2005b}  and adapt it for the
subject of this work. The main features of this code will now be
described briefly.

The reactive Euler equations are solved using an explicit piecewise
parabolic method (PPM) \citep{colella1984a}, a higher-order Godunov
scheme.  Specifically, our code is based on  the \textsc{Prometheus}
implementation \cite{fryxell1989a} of PPM.

To track the flame front we use the level-set method which was
introduced by \citet{osher1988a} and implemented in the code by
\citet{reinecke1999a}.  In this scheme, a signed distance function
$G$, which is positive in the burnt material and negative in the
unburnt material, is assigned to each point in the computational
domain. The zero level-set of $G$ thus separates the burnt from the
unburnt matter and marks the location of the flame front.  The
level-set is propagated with the burning velocity perpendicular to the
flame surface and advected as a passive scalar without fundamental
modifications of the hydrodynamics solver.  It is now possible to
calculate the burnt and unburnt volume fractions in each cell.

To ensure that the regions of highest interest are optimally resolved
for a given fixed number of grid cells and no computational resources
are wasted on regions of subordinate importance, the computational
domain is separated into two grids \citep{roepke2005c,roepke2006a},
an outer coarser grid, where the cell size increases outwards, and an
uniformly spaced moving inner grid tracking the conversion front and
expanding with it into the outer grid. This way we achieve an initial
resolution in the center of the star of $\sim 2.6\times 10^3 \,
\mathrm{cm}\times \mathrm{(grid\ cells\ per\ dimension/128)^{ -1}}$,
if our grid covers one octant of the star.

As described in Section \ref{sec:turb_comb}, we cannot resolve the
turbulent motion down to the Gibson scale.  Therefore, we perform
``large eddy simulations'': Only the largest scales of the system are
resolved, while the turbulent motion on smaller scales is modeled by a
subgrid scale (SGS) model. The SGS model determines the turbulent
energy, from which the turbulent burning velocity can be inferred.
\citet{schmidt2006b,schmidt2006c} introduced a sophisticated localised
SGS turbulence model and implemented it into the code. This model
determines the SGS turbulence velocity $q_\mathrm{SGS}$. The turbulent
burning velocity $v_\mathrm{turb}$ is then obtained by setting \citep{schmidt2006c}
\begin{eqnarray}
v_\mathrm{turb} &=& v_\mathrm{lam}
\sqrt{1+\frac{4}{3}\left(\frac{q_\mathrm{SGS}}{v_\mathrm{lam}}\right)^2},
\end{eqnarray}
with the laminar
burning velocity $v_\mathrm{lam}$ as a lower limit.
 
The code as described was written to set up a white dwarf in a
Newtonian gravitational potential and to model the thermonuclear
burning of carbon and oxygen. Thus, it had to be adapted
to the subject of this work. 
In contrast to the case of white dwarfs (compactness
$(GM/Rc^2)_{\mathrm{WD}} \sim 0.001$), in neutron stars (compactness
$(GM/Rc^2)_{\mathrm{NS}} \sim 0.1$) general relativistic effects
cannot be neglected.  Computations in full general relativity are,
however, beyond our scope.  Given the overall uncertainties,
particularly in the EoS, we consider the error introduced by the use
of Newtonian dynamics to be not critical, however a comparison of our
results with general relativistic simulations would be
interesting. But a modification of the gravitational potential cannot
be avoided, otherwise the results would be completely beside the
point. For example for a given mass of the neutron star the central
density would be much lower and thus exothermic combustion would not
be possible at all. Therefore an effective relativistic gravitational
potential \citep{marek2006a} based on the Tolman-Oppenheimer-Volkov
(TOV) equations was implemented.

Further adaptions of the code to the new setup include the EoS for
hadronic and quark matter as described in Section \ref{sec:eos}, and,
as a replacement for the ``burning routine'' in the original code, a
routine which takes care of the conversion of the hadronic matter into
SQM. This takes place by switching to the quark EoS and releasing the
difference in the energy per baryon into internal energy, while conserving
the total energy.

We set up one octant of the neutron star on a three-dimensional
Cartesian grid with 128 or 192 grid cells in each dimension and
applied reflecting boundary conditions at all borders of the
computational domain. Burning is initialized in the following way: At
the center of the star we construct a small sphere with a radius of
$r_\mathrm{seed}=10^5\,\mathrm{cm}$ on which a sinusoidal perturbation
with an amplitude of $2\times10^4\,\mathrm{cm}$ is superimposed. The
initial seed is shown in the close-up of Figure \ref{fig:150}
(a). When starting the simulation, the matter inside this small volume
is converted instantly and constitutes the initial SQM seed.

Since both the size and the form of the initial seed are not known, we
choose this configuration for numerical reasons: The size of the
perturbations is similar to the minimum length scale for turbulent
burning $\lambda_\mathrm{min}$ (cf. Sections \ref{sec:turb_comb} and
\ref{sec:turb}), therefore the front is expected to develop
Rayleigh-Taylor instabilities soon after the start of the
simulations. Smaller initial perturbations would need some time to
grow before Rayleigh-Taylor instabilities become possible. But since
in the end the core is converted completely, the results should change
only slightly, whereas the computational costs would be considerably
higher. As described in Section \ref{sec:combustion}, we assume the
combustion to be a deflagration and ignite the burning accordingly.
Although we do not expect different initial configurations to alter
our results considerably, possible effects of different initial
geometries and different ways of ignition will be explored in future
work.

\section{Simulations}\label{sec:simulations}
We conduct several runs with varying bag constant $B$. Since only
some constraints on $B$ are known, we can use it as a parameter to
change the EoS for SQM and are thus able to control the amount of
released energy from very high to rather low values,
cf. (\ref{EA}). We vary $B$ in a subset of the theoretically
admissible range between a lower limit of $B^{1/4}_{low} = 147\,
\mathrm{MeV}$ and an upper limit of $B^{1/4}_{\mathrm{high}} = 155\,
\mathrm{MeV}$.  At even higher $B$, the combustion would be restricted
to the very innermost region of the neutron star or would not be
possible at all, cf. Figure \ref{fig:plotcomb} and (\ref{exo}).  We
use $B^{1/4} > 155\, \mathrm{MeV}$ only to test if instabilities grow
at the beginning of the burning; results are presented in Section
\ref{sec:turb}.
In alternative units our chosen limits are roughly $
B_{\mathrm{low}} = 60\, \mathrm{MeV/fm^3}$ and $ B_{\mathrm{high}} =
80\, \mathrm{MeV/fm^3}$ -- values also used as a lower and upper limits
in the literature \citep[e.g.][]{bauswein2010a}.

\begin{figure}

\includegraphics[width=8.6cm]{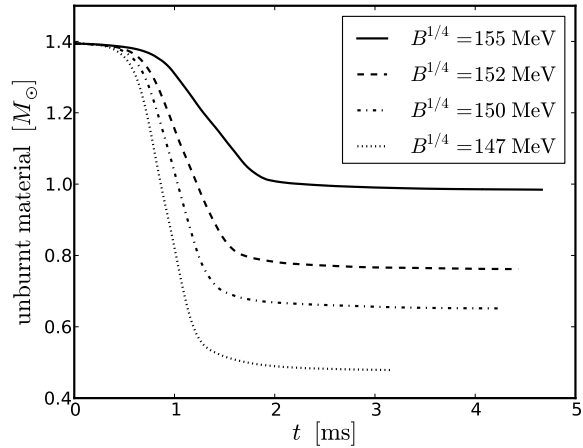}

\caption{Gravitational mass of unburnt (hadronic) material in the
  three-dimensional simulations for different bag constants $B$,
  as a function of time (models B155\_128, B152\_128, B150\_128 and B147\_128). }
\label{fig:unburnt}
\end{figure}

We start our computations with an non-rotating, cold, isothermal
``standard neutron star'' in hydrostatic equilibrium, having an
initial central total energy density of $e_c=1.0 \times 10^{15}\,
\mathrm{g/cm^3}$, a gravitational mass of $M=1.4\, M_\odot$, a radius
of $R = 11\, \mathrm{km}$, a proton fraction of $Y_p = 0.2$, and a
temperature of $T=100\, \mathrm{keV}$.  We conducted four runs
with a resolution of 128 grid cells per dimension and bag constants of
$B^{1/4} = 147\,,150\,,152\, $ and $155 \,\mathrm{MeV}$, respectively.
Table \ref{tab:overview} shows an overview of the models presented
here. In Figure \ref{fig:unburnt} the temporal evolution of the
conversion for different $B$ is shown, represented by the
gravitational mass of the remaining unburnt hadronic material.

\begin{table}
\begin{tabular}{|l|c|c|c|}
\hline
Model&Resolution&$B^{1/4}/\mathrm{MeV}$&$M_\mathrm{unburnt}/M_\odot$\\
\hline
\hline
B147\_128&$128^3$&147&0.48\\
\hline
B150\_128&$128^3$&150&0.66\\
\hline
B150\_192&$192^3$&150&0.67\\
\hline
B152\_128&$128^3$&152&0.77\\
\hline
B155\_128&$128^3$&155&0.99\\
\hline
\end{tabular}

\caption{Overview of the different models. $M_\mathrm{unburnt}$ is the
  gravitational mass of the remaining hadronic outer layer at
  $t=3.0\, \mathrm{ms}$, when the combustion can be considered as
  complete in all cases.}
\label{tab:overview}
\end{table}

\begin{figure}
\includegraphics[width=8.6cm]{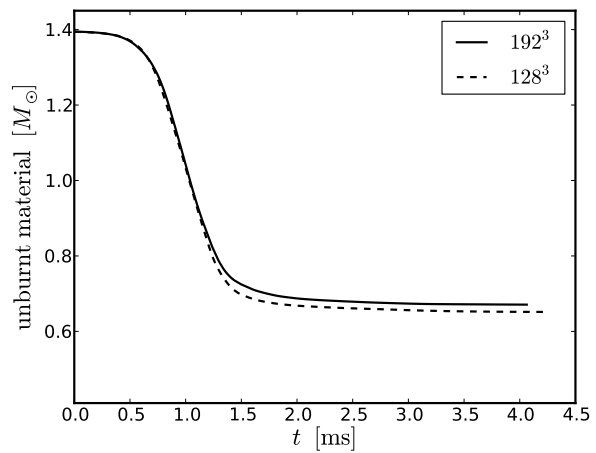}
\caption{Resolution study: Two models with $B^{1/4} =
  150\,\mathrm{MeV}$ which differ only in resolution (B150\_192 and B150\_128).}
\label{fig:resolution}
\end{figure}

In addition, we conducted one run with a higher resolution, 192 grid
cells per dimension, using an intermediate bag constant of $B^{1/4} =
150\,\mathrm{MeV}$ (model~B150\_192).  To study the effects of
different resolutions, we compare in Figure \ref{fig:resolution} the
two models B150\_128 and B150\_192, which differ only in the
resolution ($128^3$ and $192^3$, respectively).  Apparently there are
only slight differences between the two models. In particular the
slopes in the phase of rapid burning, which are determined by the
conversion rate, which in turn depends on the turbulent burning
velocity, agree very well.
The different resolutions only become noticeable in the representation
of the exact position of the density threshold for exothermic
combustion -- hence the slight discrepancy in the amount of unburnt
matter at later times.
Therefore we consider our simulations converged in the
sense that the effects caused by resolution are smaller than
uncertainties caused by other sources. Thus, we regard a resolution of
128 cells per dimension to be sufficient for our quantitative analysis.

After addressing the question of whether burning is turbulent, the
results of the simulation with the highest resolution, model
B150\_192, are discussed in some detail below.  In the subsequent
sections we will briefly discuss differences in the two extreme cases
(models B147\_128 and B155\_128).

\subsection{Onset of Turbulence}\label{sec:turb}
\begin{table}
\begin{tabular}{|l||c|c|c|c|c|c|}
\hline
$B^{1/4}/\mathrm{MeV}$&145&147&150&152&155&157\\
\hline
$At$&0.11&0.091&0.067&0.051&0.027&0.010\\
\hline
$\lambda_\mathrm{min}/10^4\,\mathrm{cm}$&$3.6$&$4.4$&$6.2$&$8.5$&$16$&$45$\\
\hline

\end{tabular}

\caption{Atwood number $At$ and minimal length scale for turbulent
  burning $\lambda_\mathrm{min}$ for different bag constants $B$ at
  time $t = 0.1\,\mathrm{ms}$, determined in three-dimensional
  simulations.}
\label{tab:atwood}
\end{table}

\begin{figure}
\includegraphics[width=8.6cm]{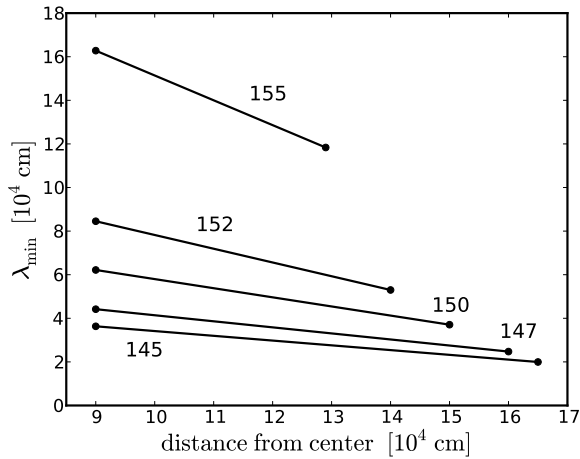}

\caption{Comparison of the minimal length scale for turbulent burning
  $\lambda_\mathrm{min}$ in the early phase of the conversion process
  for different bag constants $B$ and points in time, determined in
  three-dimensional simulations. The number on each line indicates
  $B^{1/4}$ in $MeV$. For each $B$ the first and second point
  correspond to time $t_1 = 0.1\, \mathrm{ms}$ and $t_2 = 0.2\,
  \mathrm{ms}$, respectively. On the abscissa the average position of
  the conversion front at $t_1$ and $t_2$ is shown. }
\label{fig:lambdamin}
\end{figure}

We calculate the minimum length scale for turbulent burning,
$\lambda_\mathrm{min}$, according to (\ref{eq:lambdamin}), see Section
\ref{sec:turb_comb}. To ensure comparability, we use  the
same three-dimensional setup for all $B$, as described above, and the same
resolution of 128 grid cells in each dimension.

In Figure \ref{fig:lambdamin} we compare $\lambda_\mathrm{min}$ at the
beginning of the conversion process for different bag constants $B$
and points in time.
The density contrast is quantified by the \textit{Atwood number}
$At=(e_h-e_q)/(e_h+e_q)$. Table \ref{tab:atwood} lists $At$ and
$\lambda_\mathrm{min}$ for different $B$. The values were
determined at $t = 0.1\,\mathrm{ms}$.
As visible in Table \ref{tab:atwood} and Figure \ref{fig:lambdamin},
$\lambda_\mathrm{min}$ depends strongly on $B$, and becomes very large
for high $B$.  For the highest examined bag constant, $B^{1/4} = 157\,
\mathrm{MeV}$, $\lambda_\mathrm{min}$ is comparable to the size of the
system and no growth of Rayleigh-Taylor instabilities is expected.
Bag constants starting at $B^{1/4} = 152\, \mathrm{MeV}$ down to the
lowest $B$, lead to smaller $\lambda_\mathrm{min}$ which are
comparable to or smaller than the size of the initial perturbations --
thus instabilities can grow. In the simulation with $B^{1/4} = 155\,
\mathrm{MeV}$ this is not the case at $t = 0.1\,\mathrm{ms}$ but
already at some slightly later time, since $\lambda_\mathrm{min}$
decreases with time as the gravitational acceleration becomes
stronger, cf. (\ref{eq:lambdamin}) and Figure \ref{fig:lambdamin}. Our
simulations confirm this: in all runs except for $B^{1/4} \gtrsim
157\, \mathrm{MeV}$ we see Rayleigh-Taylor instabilities form. Thus
the burning of a neutron star into a quark star becomes turbulent in
most cases, given our choice of EoS.

\subsection{Intermediate Case: $B^{1/4} = 150\, \mathrm{MeV}$}

\begin{figure}
\includegraphics[width=8.6cm]{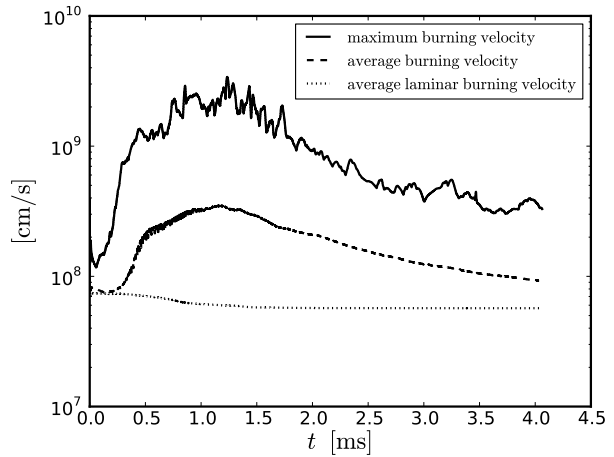}

\caption{Burning velocity: Comparison at each timestep of maximum
  burning velocity, average burning velocity and the underlying
  average laminar burning velocity. The averages where done over all
  cells in which burning occurs. Data from the high resolution run
  with $ B^{1/4} = 150\, \mathrm{MeV}$ (model B150\_192).}
\label{fig:vburn150}
\end{figure}

\begin{figure*}
  \centering
  \begin{minipage}[b]{0.45\linewidth}\centering
    \includegraphics[width=\linewidth]{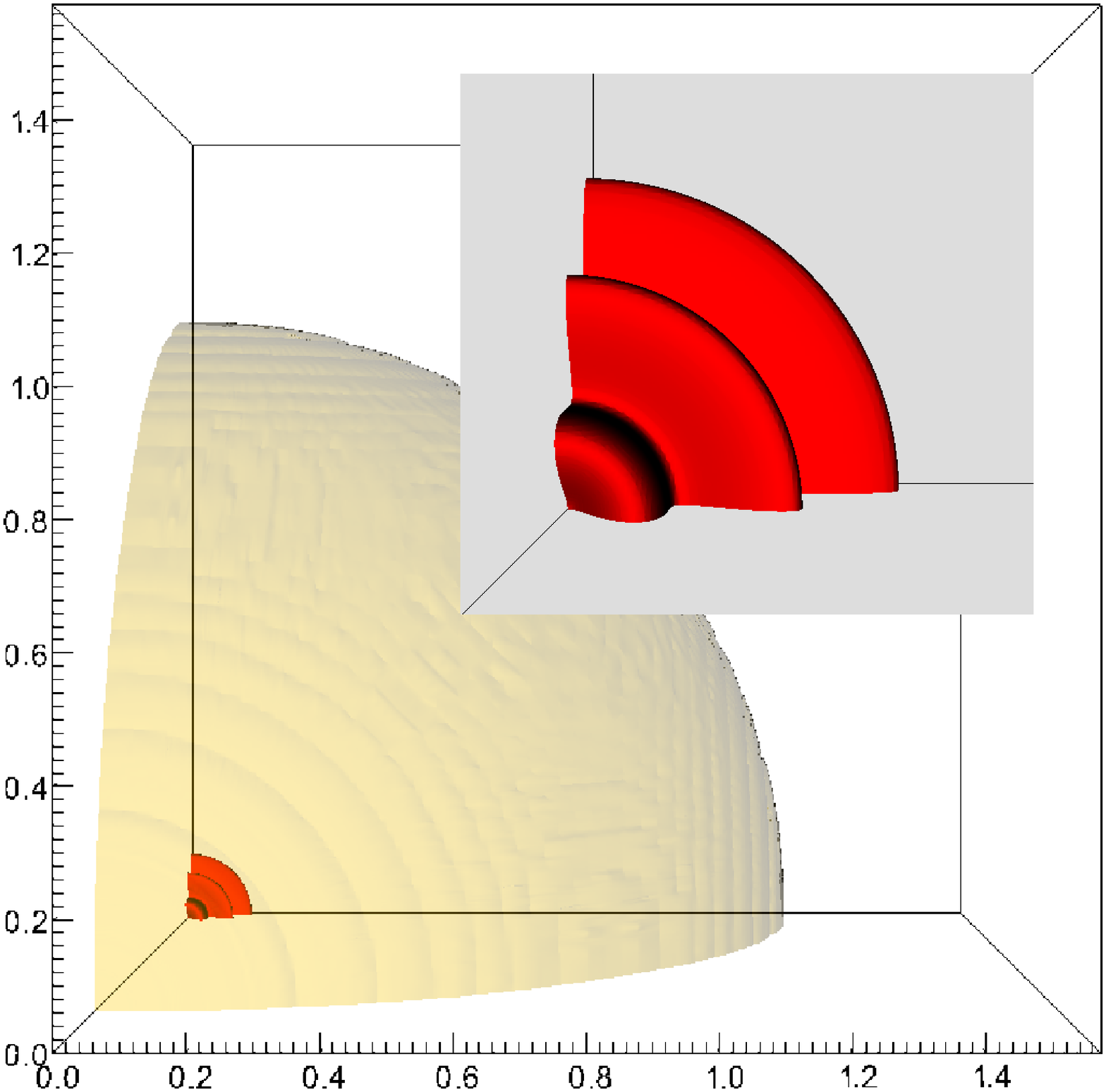}\\
    (a) $t = 0$
  \end{minipage}
  \hspace{0.05\linewidth}
  \begin{minipage}[b]{0.45\linewidth}\centering
    \includegraphics[width=\linewidth]{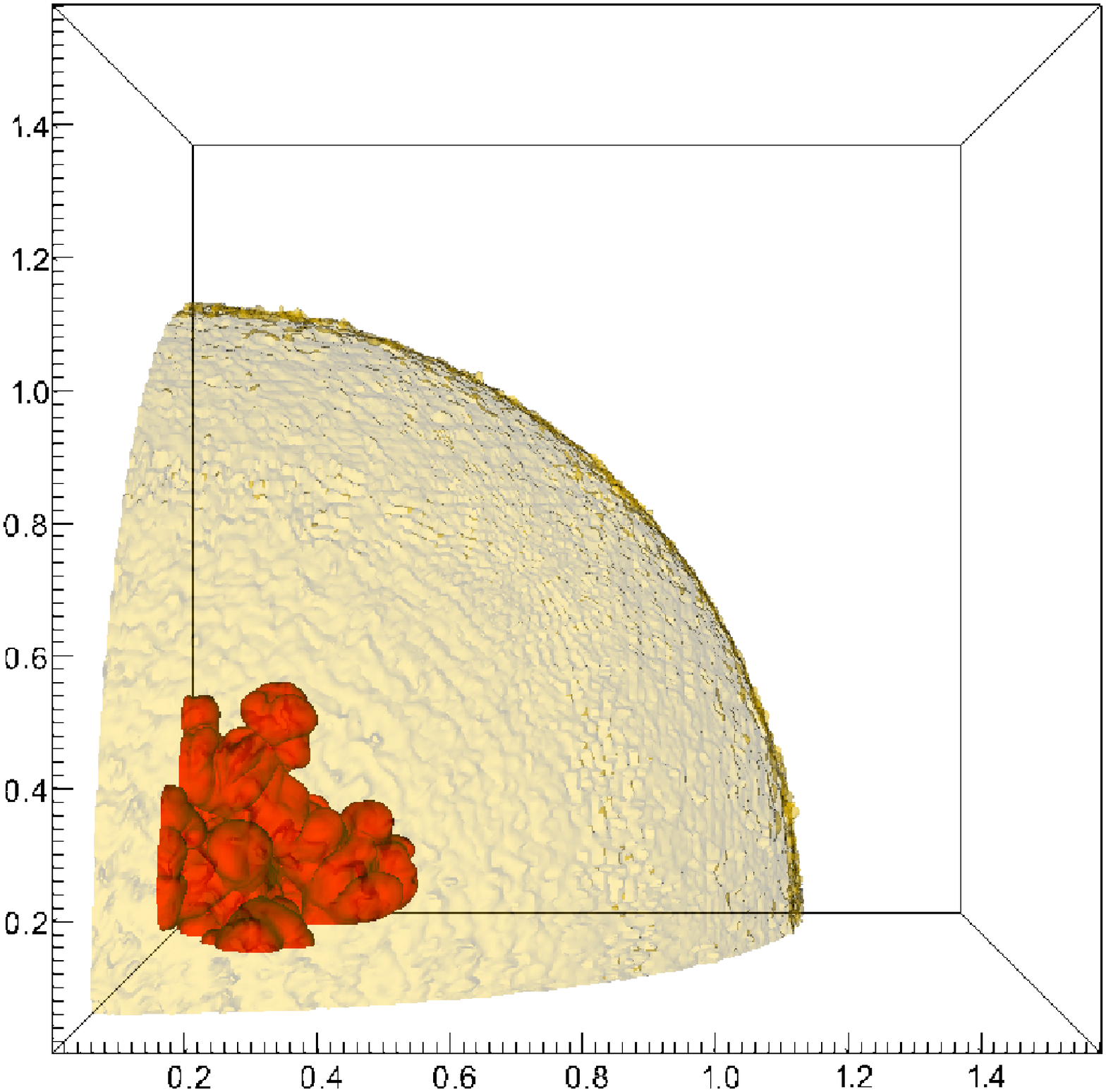}\\
    (b) $t = 0.7\, \mathrm{ms}$
  \end{minipage}

  \begin{minipage}[b]{0.45\linewidth}\centering
    \includegraphics[width=\linewidth]{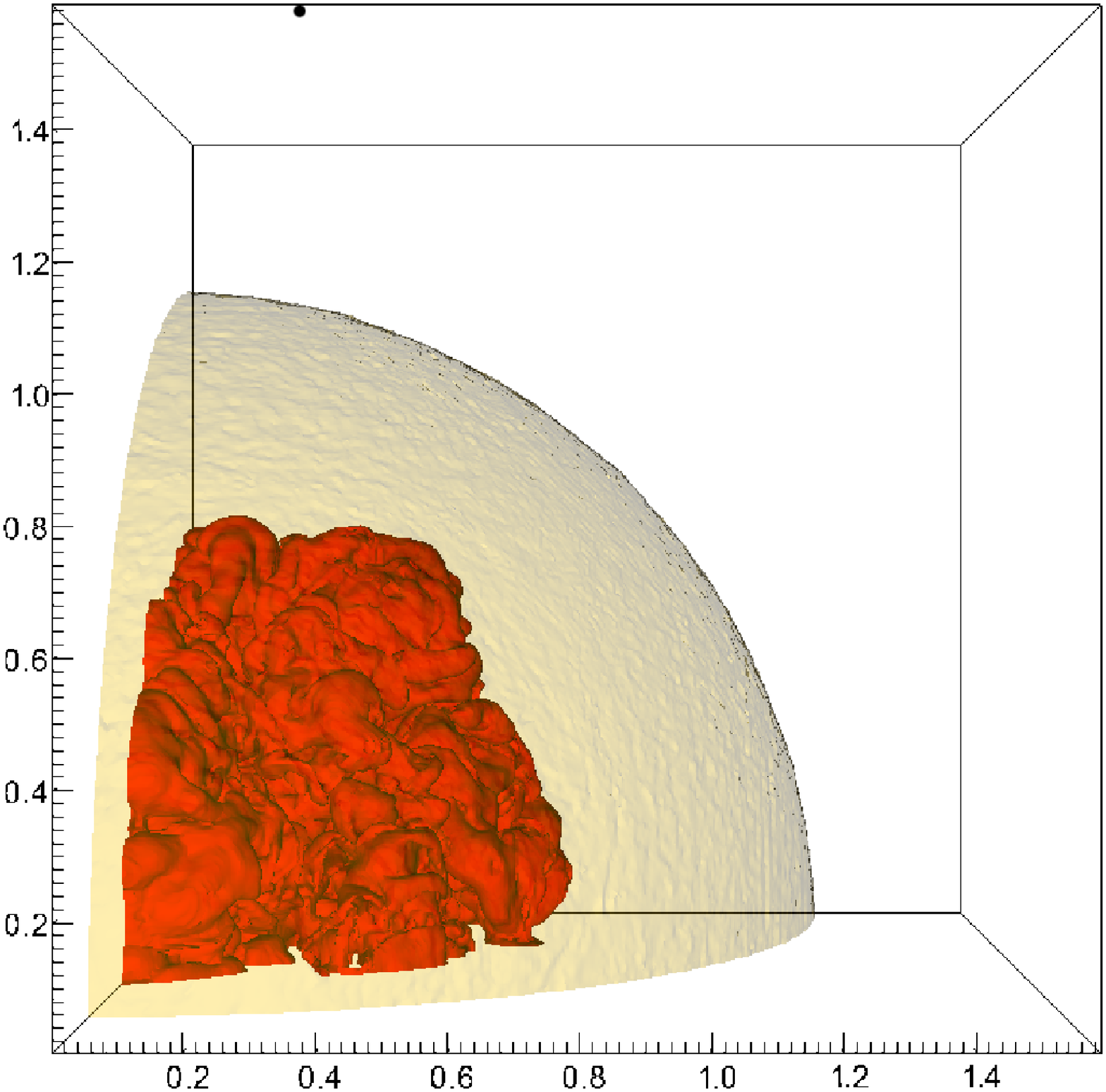}\\
    (c) $t = 1.2\, \mathrm{ms}$
  \end{minipage}
  \hspace{0.05\linewidth}
  \begin{minipage}[b]{0.45\linewidth}\centering
    \includegraphics[width=\linewidth]{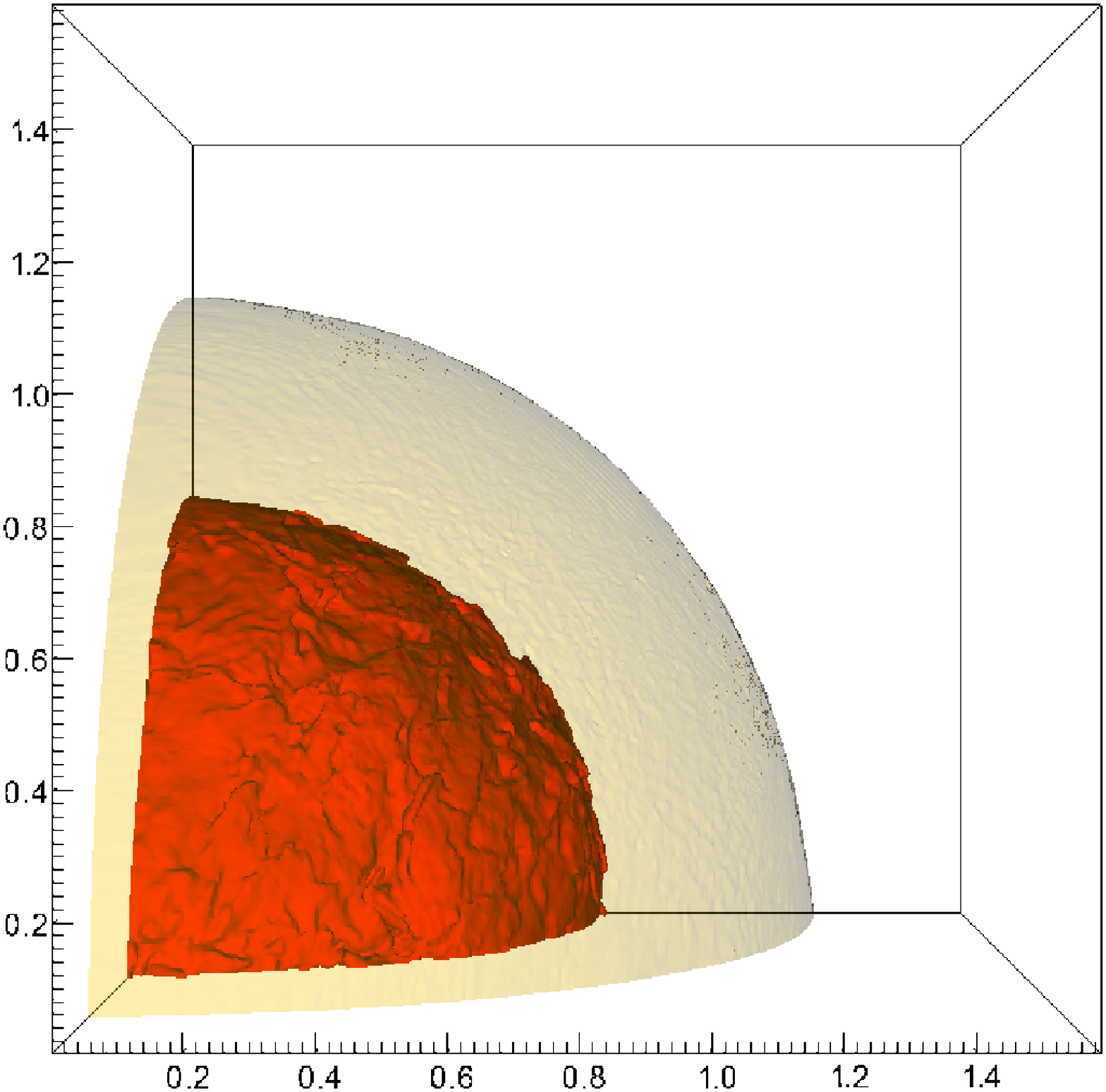}\\
    (d) $t = 4.0\, \mathrm{ms}$
  \end{minipage}

  \caption{(color online) Model B150\_192: Conversion front (red) and
    surface of the neutron star (yellow) at different times $t$. In
    (a) a close-up of the central region is added. Spatial units
    $10^6\,\mathrm{cm}$. }
  \label{fig:150}
\end{figure*}

In this section we present a detailed discussion of the results of the
simulation with a resolution of 192 grid cells per dimension and an
intermediate bag constant, $B^{1/4} = 150\, \mathrm{MeV}$ (model
B150\_192).  According to (\ref{EA}) the energy per baryon in this
case is $E/A = 858\, \mathrm{MeV}$, corresponding to a difference of
$\sim70\, \mathrm{MeV}$ per baryon with respect to the energy of
nuclear matter.

In Figure \ref{fig:150} (a) the initial configuration including the
SQM seed in the center can be seen. The shape of the seed as described
in Section \ref{sec:numerical_method} is shown additionally in the
close-up in this figure.
After ignition the conversion front propagates into the hadronic
matter, at first in a laminar way until initial perturbations of the
conversion front become unstable due to Rayleigh-Taylor
instabilities. Until turbulence has fully developed, the conversion
process stays in a short phase of nearly laminar burning while the
instabilities grow, see Figure \ref{fig:unburnt} which shows the
amount of unburnt (hadronic) matter as a function of time, and Figure
\ref{fig:vburn150}, where we compare the average laminar burning
velocity, the average burning velocity and the maximum burning
velocity at each timestep. The averaging was done over all cells in
which burning occurs.

As the instabilities grow, typical mushroom-shaped structures, rising
plumes of SQM, are forming and hadronic matter is falling down in
between. These structures can be seen in Figure \ref{fig:150} (b),
where the conversion surface is shown at $t=0.7\,
\mathrm{ms}$. Starting at $t\sim0.5\, \mathrm{ms}$ strong turbulence
develops and rapid burning takes place until $t\sim1.5\, \mathrm{ms}$,
as visible in Figure \ref{fig:unburnt}. The structure of the
conversion front near the end of this phase of rapid burning can be
seen in Figure \ref{fig:150} (c). The plumes grow until the conversion
front reaches densities where the condition for exothermic combustion
(\ref{exo}) is no longer fulfilled. They continue to grow laterally,
until they eventually merge, leaving bubbles of hadronic matter in
between. Turbulence then weakens and the flame slows down.  The
remaining pockets filled with hadronic material shrink until they
eventually vanish completely. Now all matter inside the volume
confined by the above mentioned density threshold is burned and the
star consists of an inner sphere of SQM containing about half of the
mass and an outer layer of unburnt hadronic matter (cf. Figure
\ref{fig:150} (d)). This outer layer has a mass of about $0.67\,
\mathrm{M_\odot}$ and densities lower than the threshold (\ref{exo})
but mostly still super-nuclear (applying LS EoS and $ B^{1/4} = 150\,
\mathrm{MeV}$, the density threshold in cold matter is at about $1.8\,
\rho_{\mathrm{nuclear}}$, see Figure \ref{fig:plotcomb2}).

Turbulent motions leads to burning velocities considerably higher
than the laminar burning velocities, the maximum amplification factor
is about 50 and on average between 2 and 20 (see Figure
\ref{fig:vburn150}). This figure also clearly shows  that the turbulent
burning velocity and thus the strength of the turbulence increases
rapidly until it reaches a maximum at $t \sim 1.0\,
\mathrm{ms}$. At that point a steady but slower decrease starts.
The maximum Mach numbers reached were about 0.2, i.e.\ the
combustion was clearly subsonic.
As we do not include any kind of cooling, the large amount of energy
released in the burning process is turned into thermal energy and the
inner SQM region is heated to temperatures of about $50\,
\mathrm{MeV}$ in the center of the star.

We stopped this simulation at $t = 4.0\, \mathrm{ms}$. By then the
conversion rate has dropped to a very low value and seems to approach
zero asymptotically. Since at that time the system is approximately in
hydrostatic equilibrium (the dynamical time scale of a neutron star is
$\tau_\mathrm{dyn}\sim5\times 10^{-2}\,\mathrm{ms}$) we do not
expect any further conversion of a significant amount of
mass. Therefore the structure of the remnant should not change if the
simulation would have been carried on for longer timescales -- at
least in our model without cooling processes and in the approximation
of a hydrodynamic combustion.

\subsection{Lower Limit: $B^{1/4}_\mathrm{low} = 147\, \mathrm{MeV}$}\label{sec:lowbag}

\begin{figure*}
  \centering
  \begin{minipage}[b]{0.45\linewidth}\centering
    \includegraphics[width=\linewidth]{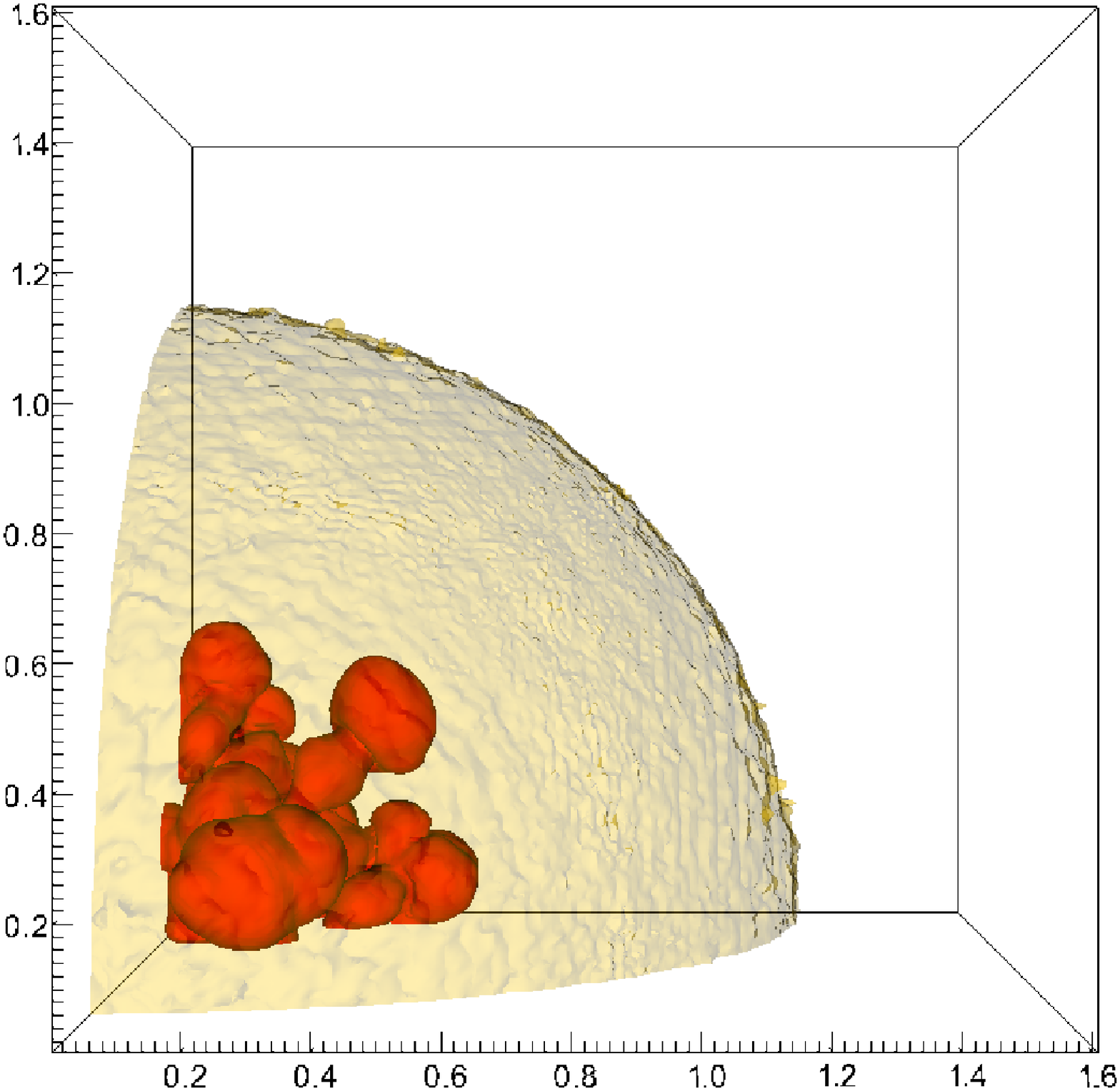}\\
    (a) $t = 0.7\, \mathrm{ms}$
  \end{minipage}
  \hspace{0.05\linewidth}
  \begin{minipage}[b]{0.45\linewidth}\centering
    \includegraphics[width=\linewidth]{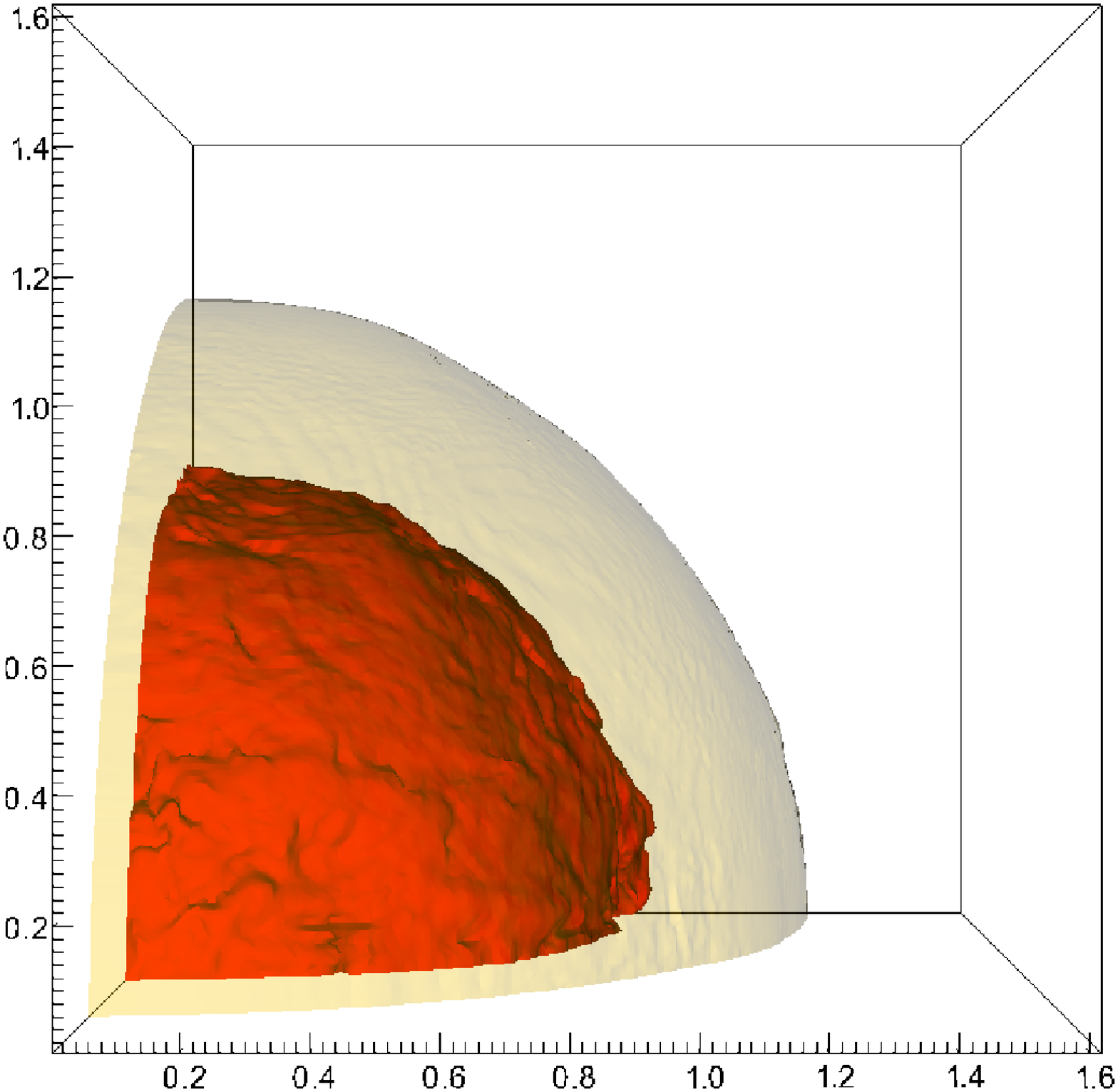}\\
    (b) $t = 3.1\, \mathrm{ms}$
  \end{minipage}
\caption{(color online) Model B147\_128: Conversion front (red) and
    surface of the neutron star (yellow) at different times $t$. Spatial
    units $10^6\,\mathrm{cm}$.}
\label{fig:147}
\end{figure*}

Now we briefly discuss the simulation with 128 grid cells per dimension
and with our lower limit for the bag constant, $
B^{1/4}_{\mathrm{low}} = 147\, \mathrm{MeV}$ (model B147\_128).  This
corresponds to the largest difference in energy per baryon
compared to nuclear matter, $E/A = 90\, \mathrm{MeV}$.

Qualitatively, the conversion process evolves in the same way as in the
case described above (model B150\_192), but there are some
quantitative differences: The energy release is higher than in the
intermediate case, therefore the burning leads to a stronger inverse
density stratification, resulting in a faster growth of instabilities
and stronger turbulence. The rising plumes of SQM are observable in
Figure \ref{fig:147} (a) as typical ``mushrooms'', like in the
previous case. Comparing Figure \ref{fig:147} (a) and Figure
\ref{fig:150} (b), both showing the conversion surface at $t = 0.7\,
\mathrm{ms}$, clarifies that the conversion process takes place
considerably faster for the lower $B$. Figure \ref{fig:unburnt} shows
that after a short phase of slow burning, rapid burning occurs from
$t\sim 0.4\, \mathrm{ms}$ until $t\sim1.5\,\mathrm{ms}$. Then the
burning slows down and the conversion rate approaches zero. At $t=5\,
\mathrm{ms}$, the remnant has an inner SQM core with a radius of $\sim
9 \, \mathrm{km}$, cf.\ Figure \ref{fig:147} (b), surrounded by an
hadronic outer layer with a mass of $0.48 \, \mathrm{M_\odot}$, the
least massive outer layer in all our simulations.
Central temperatures of the core reach $53\, \mathrm{MeV}$,
somewhat higher than in the previous case due to the higher energy
release.

\subsection{Upper Limit: $B^{1/4}_\mathrm{high} = 155\, \mathrm{MeV}$}\label{sec:hibag}

\begin{figure*}
  \centering
  \begin{minipage}[b]{0.45\linewidth}\centering
    \includegraphics[width=\linewidth]{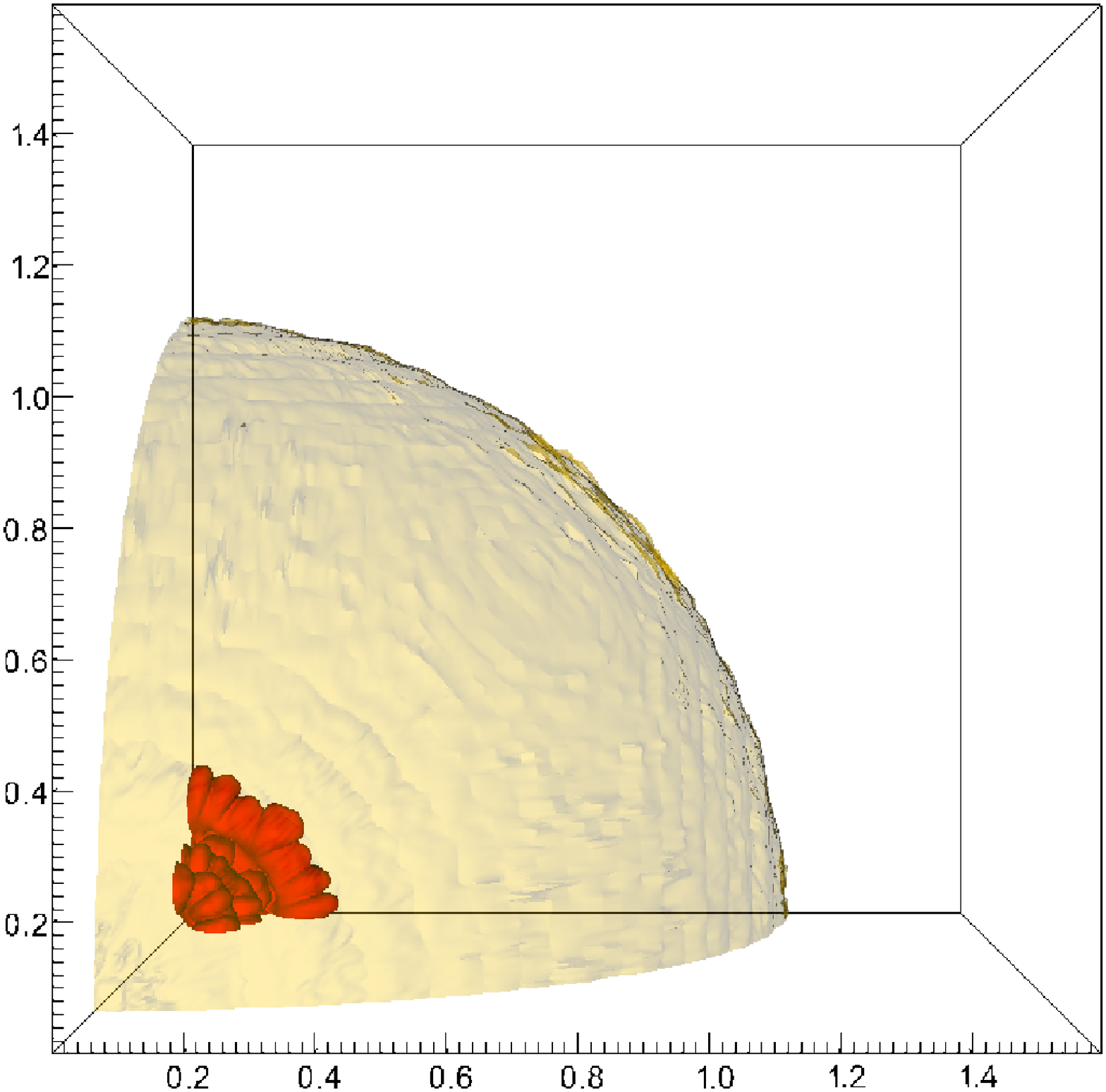}\\
    (a) $t = 0.7\, \mathrm{ms}$
  \end{minipage}
  \hspace{0.05\linewidth}
  \begin{minipage}[b]{0.45\linewidth}\centering
    \includegraphics[width=\linewidth]{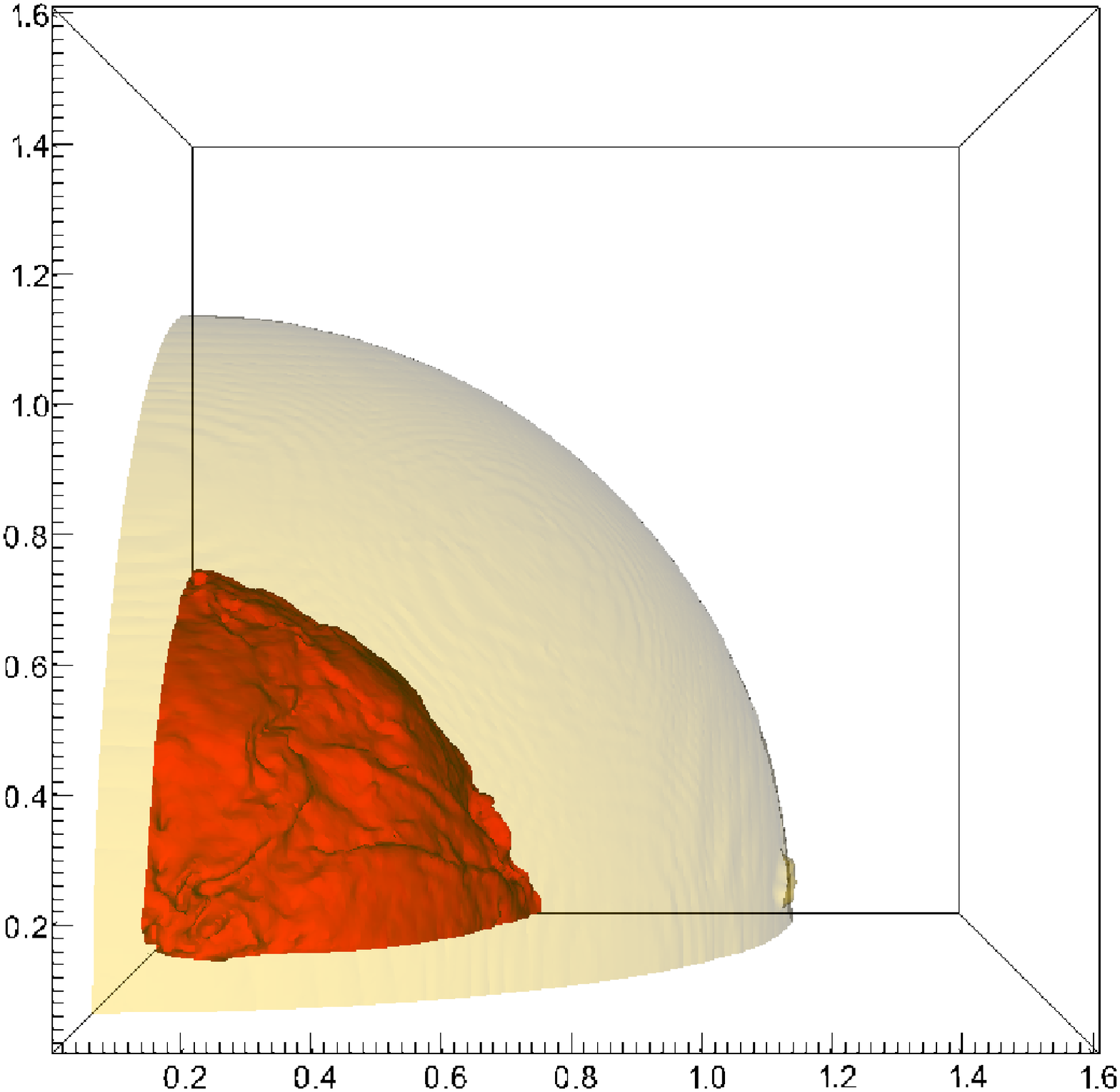}\\
    (b) $t = 4.6\, \mathrm{ms}$
  \end{minipage}
  \caption{(color online) Model B155\_128: Conversion front (red) and
    surface of the neutron star (yellow) at different times
    $t$. Spatial units $10^6\,\mathrm{cm}$.}
\label{fig:155}
\end{figure*}

Finally we present the simulation with 128 grid cells per dimension
and our highest bag constant, $B^{1/4}_{\mathrm{high}} = 155\,
\mathrm{MeV}$ (model B155\_128).  Here the difference in energy per
baryon, $E/A \sim 40\, \mathrm{MeV}$, is considerably lower than in
the cases B147\_128 and B150\_128.  Figures \ref{fig:155} (a) and
\ref{fig:155} (b) show the conversion front at $t=0.7\,\mathrm{ms}$
and at the point when we stopped our simulation, at
$t=4.6\,\mathrm{ms}$. From the figures the similar evolution compared
to the above described cases with lower $B$ are visible.  The lower
$E/A$ and the higher density threshold for exothermic burning
(cf. Figure \ref{fig:plotcomb2}) lead to a slower and less violent
burning, which ceases at higher densities compared to the models
previously shown. Consequently, at the end of the simulation the
resulting strange matter core is smaller and is surrounded by an
hadronic outer layer of $0.98 \, \mathrm{M_\odot}$.  Temperatures of
around $45\, \mathrm{MeV}$ are reached in the center. Figure
\ref{fig:unburnt} shows that the conversion rate, represented by the
slope of the curves, is lower than in the other cases and the
combustion takes longer although less material is burnt.

\section{Conclusions}\label{sec:conclusions}
We presented three-dimensional hydrodynamic simulations of the
conversion of a neutron star into a quark star assuming different bag
constants $B$ for describing SQM. In all cases we observe
growing Rayleigh-Taylor instabilities of the conversion front. The
resulting turbulent motion enhances the conversion velocity strongly,
leading to conversion timescales of $\tau_\mathrm{burn}\sim 2\,
\mathrm{ms}$ for all $B$.
However, recent suggestions \citep{niebergal2010a,horvath2010a} that
the turbulence enhances the burning speed to sonic or even supersonic
velocities could not be confirmed, which came as no surprise since in
the analogous case of SN~Ia such a transition is not possible either
as long as burning proceeds in the flamelet regime
\citep{niemeyer1997b}.

In all cases we observe at the end of our simulations a spherical SQM
interior surrounded by an outer layer of hadronic matter.
This outer layer exists because in our hydrodynamic approximation
the combustion stops when the conversion front reaches conditions
under which exothermic burning is no longer possible.
Since this condition depends on density and is fulfilled for
sufficiently high densities only, it can roughly be described as a density
threshold which forms a boundary that separates the high density
(burnt) strange quark matter and the low density (unburnt) hadronic
matter.
In our approximation we can make no statement on whether the conversion
process proceeds further beyond this boundary by processes which
cannot be described as a combustion. Possibly free neutrons diffuse
into the quark matter and are converted subsequently
\citep{olinto1987a}, a process that probably is exothermic, as
\citet{lugones1994a} already pointed out. Free neutrons are abundant
in hadronic matter at densities higher than the neutron drip density,
$e_\mathrm{drip}\sim 4\times 10^{11}\,\mathrm{g/cm^3}$. However we
expect these additional processes to happen on much longer timescales
than the combustion described in this work.

The obvious consequence of an at least temporary existence of an outer
layer of unburnt hadronic matter is that the resulting quark star
could support a rather thick crust, unlike bare strange stars, which
can presumably support only a tiny crust. This would allow e.g.\ for
pulsar glitches, if the timescale of the conversion after the
combustion has ceased is large enough.
Some authors suggested that the conversion of a neutron star into a
strange star may eject neutron-rich material from the surface, and
that in this ejecta the nucleosynthesis of heavy neutron-rich nuclei
via the r-process may occur \citep{jaikumar2007a}. However, our
results suggest that ejection of matter from the star is rather
unlikely since the violent burning ceases before reaching the
surface. Any subsequent continuation of the conversion by processes
not describable by a combustion is expected to be much slower, and to
take place in a much less violent way. But given our ignorance about
these processes more detailed work on this subject may lead to
differing conclusions.

The existence of the hadronic outer layers, or the possibility of
exothermic combustion even in the center of neutron stars, depends
(like many other properties) strongly on the EoS used for the hadronic as
well as for the quark phase. Hence any firm prediction needs
a more realistic treatment.
Furthermore, the maximum mass configuration of non-rotating stars of
both the LS EoS and our bag model EoS have $M_\mathrm{max}<2\,
M_\odot$ and therefore conflict with observations
\citep{demorest2010a}.  As mentioned before, we nevertheless use those
EoS in this work because we consider them as sufficient for our first
attempts.
In future work we want to improve on this and plan to use more
realistic EoS.  Regarding the quark phase, finite strange quark masses
and QCD-interactions can be included into the bag model.
SQM bag model EoS which contain these corrections can support a $2\,
M_\odot$ neutron star, as was shown by \citet{weissenborn2011a}.
Recently also the choice of micro-physical finite temperature EoS for
nuclear matter has become larger \citep[e.g.][]{hempel2010a,
  typel2010a}, so we can consider additional hadronic EoS apart from
the LS and Shen EoS which we used in this work.  Another possibility
is to consider the use of modern zero-temperature micro-physical EoS
together with an ideal gas component to account for temperature
effects, whose reliability has been tested in
\citep{bauswein2010b}. Further improvement would be achieved by adding
neutrino cooling, which could be relevant since rather high
temperatures are reached in the quark core.  Until now we use an
initial model resembling an old isolated neutron star, the same
calculations could be done with a young (proto)neutron star and in
connection with a core collapse supernova.
Furthermore our computations should be extended to make statements
about observable quantities. Therefore we plan to calculate the
gravitational wave signal of the conversion of a neutron star into a
quark star.

\acknowledgments{The computations for this work have been carried out
  at the Rechenzentrum Garching of the Max Planck Society.
We thank A.~Bauswein, D.~Blaschke, T.~Fischer, G.~Pagliara, M.~Hempel,
and J.~Schaffner-Bielich for helpful discussions.
This work was supported by \mbox{CompStar}, a Research Networking Programme
of the European Science Foundation. The work of FR has been supported
by Deutsche Forschungs\-gemeinschaft via the Transregional
Collaborative Research Center TRR 33 ``The Dark Universe'', the
Excellence Cluster EXC153 ``Origin and Structure of the Universe'' and
the Emmy Noether Program (RO 3676/1-1).

\end{document}